\documentclass[doublecol]{epl2}

\usepackage{graphicx}
\usepackage{latexsym} 
\usepackage[centertags]{amsmath} 
\usepackage{amsfonts} 
\usepackage{amssymb} 
\usepackage{amsthm} 
\usepackage{newlfont} 
\usepackage{subfigure} 
\usepackage{color}

\title{Identification of the critical temperature from non-equilibrium 
time-dependent quantities}

\author{E. Lippiello \inst{1} \and A. Sarracino \inst{2,3}} 
\institute{
\inst{1}Dipartimento di Scienze Ambientali and CNISM, Seconda Universit\`a di Napoli,  
Via Vivaldi, 81100 Caserta (CE), Italy \\ 
\inst{2} Dipartimento di Matematica ed Informatica, 
Universit\`a di Salerno, Via Ponte don Melillo, 84084 Fisciano (SA), Italy \\
\inst{3} CNR-ISC and Dipartimento di Fisica, Universit\`a Sapienza, p.le A. Moro 2, 00185 Roma, Italy} 

\abstract{ 
We present a new procedure that can identify and measure the critical temperature. 
This method is based on the divergence of the 
relaxation time  approaching the critical point in quenches from infinite 
temperature. We introduce a dimensionless quantity 
that turns out to be time-independent at the critical temperature. 
The procedure does not need equilibration and allows 
for a relatively fast identification of the critical temperature. 
The method is first tested in the ferromagnetic Ising 
model and in the two dimensional EA model and then applied to the one-dimensional Ising spin glass with power-law interactions.  
Here we always find a finite critical temperature also in presence of a uniform external field,
in agreement with the mean-field picture for the low temperature phase of spin glasses.}

\begin{document} 

\maketitle

\def\be{\begin{equation}} 
\def\ee{\end{equation}} 

\def\be{\begin{equation}} 
\def\ee{\end{equation}} 
\def\bfi{\begin{figure}} 
\def\efi{\end{figure}} 
\def\bea{\begin{eqnarray}} 
\def\eea{\end{eqnarray}} 
\newcommand{\ket}[1]{\vert#1\rangle} 
\newcommand{\bra}[1]{\langle#1\vert} 
\newcommand{\braket}[2]{\langle #1 \vert #2 \rangle} 
\newcommand{\ketbra}[2]{\vert #1 \rangle  \langle #2 \vert}

The identification  of a critical temperature $T_c$ is a 
fundamental characterization of statistical systems. 
This, indeed, allows one to construct the phase-diagram of the  
system and to obtain insights in the underlying relevant physical 
mechanisms. In many cases the existence/absence of a phase-transition 
discriminates among different pictures for a given system.   
A major example is the long-standing question on the nature of the 
spin-glass phase in finite dimensional systems.     
According to the replica symmetry breaking  scenario \cite{parisi}, a transition line should exist,  
the so-called de Almeida-Thouless (AT) line,
separating the paramagnetic from the spin glass phase in the 
temperature \emph{vs} magnetic field phase diagram~\cite{almeida,parisi}.  
More precisely, the spin glass phase is not destroyed by applying an external field. 
Conversely, other theories, as the ``droplet picture''~\cite{fisher.huse}, predict 
no AT line. Therefore, the identification of a 
critical temperature in the presence of an external perturbation 
discriminates between the different theories. 

The critical point  is usually identified with the temperature $T=T_c$  where the  
equilibrium correlation length $\xi(T)$ diverges. This property reflects on the behavior 
of the order parameter correlation function $C(r)$ whose asymptotic decay changes from 
exponential to algebraic when approaching $T_c$. Such a 
study is often hindered by the fact that also the relaxation 
time $t_{eq}$ diverges for temperatures close to $T_c$ and it is only possible 
to equilibrate systems of small size $L$.  
However, it is possible  
to extrapolate the critical temperature in the $L \to \infty$ 
thermodynamic limit from the behavior of finite systems
   by means of well established methods such as 
Finite Size Scaling (FSS)~\cite{fss}.  The most used method for 
the identification of $T_c$, generally
known as Phenomenological Renormalization,   
consists in the 
introduction of an appropriate dimensionless quantity that is expected
to cross exactly at the critical point \cite{nightingale}.  
In the specific case of magnetic systems, one usually measures \cite{Ballesteros1,Ballesteros2,Luescher,Palassini} 
a finite size correlation length  $\xi(L,T)$ at the temperature 
$T$, as for instance $\xi(L,T)^2=\left(\int_0^L dr~r^2C(r,T)\right)/\int_0^LdrC(r,T)$~\cite{cooper}. 
The dimensionless quantity $X(L,T)=\xi(L,T)/L$ is then plotted 
over $T$ for different values of $L$. According to FSS, one 
expects the following relation 
\be 
X(L,T)=f\left[\xi(T)/L \right] , 
\label{FSS} 
\ee 
implying that  $X(L,T)$ becomes $L$ independent when $\xi(T)$ 
diverges, namely for $T=T_c$. $T_c$ is therefore  given 
by the temperature where curves for different $L$ intersect.

On the other hand one can identify $T_c$ exploiting the intrinsically 
non-equilibrium nature of the critical point as, for instance, 
implemented in Short Time Dynamics (STD)~\cite{std},  
Non-Equilibrium Relaxation (NER)~\cite{ner} or other 
methods which use both equilibrium 
and non-equilibrium measurements~\cite{campbell}.  
These are substantially based on the dynamical scaling 
hypothesis assuming  the existence at the time $t$ of a  typical 
dynamic length $L(t)$. The divergence of the relaxation time at $T_c$ 
then reflects on a power law temporal decay of  
some observables such as the magnetization.  

In this letter, exploiting dynamical scaling as in NER methods, we propose a 
procedure to identify $T_c$ that uses phenomenological renormalization
techniques with non-equilibrium time 
dependent quantities. More precisely, we set $L\to \infty$ from the 
beginning, and replace in Eq.~(\ref{FSS}) space distances with 
times introducing the dimensionless quantity 
\be 
X(\tau,T)=\frac{1}{\tau} \frac{\int _0^\tau dt~t C(t,T)}{\int _0^\tau dt C(t,T)}, 
\label{FTS} 
\ee 
where $t$ is the time since the quench at the temperature $T$ from a 
disordered  initial condition, $\tau$ is a fixed time
and $C(t,T)$ is the correlation function with the initial 
configuration. For instance, for spin systems, $C(t,T)=\langle 
(s_i(t)-\langle s_i(t) \rangle) (s_i(0)-\langle 
s_i(0) \rangle) \rangle$, with $s_i(t)$ the spin in the 
position $i$ at time $t$. The average is performed 
over initial conditions, 
non-equilibrium dynamics at the temperature $T$, and quenched disorder
if present. We always take initial conditions corresponding to
equilibrium at infinite temperature, and therefore $\langle
s_i(0)\rangle=0$ and $C(t,T)=\langle s_i(t)s_i(0)\rangle$.   
Let us notice that other quantities can be 
equivalently considered  in the definition of $X(\tau,T)$. Here we use 
the two-time correlation function, that is more easily obtained in 
numerical simulations. However  one can also use the thermo-remanent 
magnetization, more suitable in  experimental settings.    
Dynamical scaling, then, predicts 
\be 
X(\tau,T)=g\left[t_{eq}(T)/\tau \right] , 
\label{FTS2} 
\ee  
where $t_{eq}(T)$ is the relaxation time for quenches at the 
temperature $T$. The divergence of $t_{eq}(T)$ approaching $T_c$ 
implies that $X(\tau,T)$ does not depend on $\tau$ when $T=T_c$.  
Therefore, considering different values of $\tau$ and plotting 
$X(\tau,T)$ vs $T$, one identifies $T_c$ as the intersection point of 
the different curves.  

The above procedure does not require to equilibrate the system, 
leading to some advantages with respect to the FSS method. 
Indeed, in experiments, the sample sizes always fulfill the thermodynamic limit and 
one cannot explore the $L$ dependence. Conversely,
one can measure $X(\tau,T)$ for different time interval $\tau$ and different $T$.
From a numerical point of view, in particular for spin-glasses, the 
equilibration of the system is numerically hard to
  achieve and to check. Indeed, equilibration is very time-demanding and 
indirect checks are always necessary in order to verify that a true 
equilibrium state is attained. 
Moreover, in our approach, by means of a single simulation 
up to the final time $t_f$, one can compute $X(\tau,T)$ 
for many values of $\tau \in [0,t_f]$. Conversely, in the FSS method 
one has just one $X(L,T)$ for each simulation with a system of size $L$.   
Another difference relies on the possibility of using simple
spin-spin correlations as $C(t,T)$ also 
in disordered systems, where, generally, the
  identification of a correlation length
necessitates the computation of multi-spin 
correlation functions~\cite{franz}. The presence of disorder, indeed, makes
 $\xi(L,T)$ invisible to equal time  spin-spin correlation
 function. Conversely, $C(t,T)$ is strongly affected by $t_{eq}(T)$.
 A final remark concerns the possibility to better control the influence
 of scaling corrections in the estimated $T_c$. Indeed, a pure power
 law decay  $C(t,T)\sim t^{-\theta_c}$ must be observed at
 $T=T_c$, leading to  $X(\tau,T_c)=(1-\theta_c)/(2-\theta_c)$.
The exponent $\theta$ is related to the Fisher-Huse
  exponent $\lambda$ via the relation $\theta=\lambda/z$, where $z$ is
  the growth exponent.

The measured $X$ at the intersection point, therefore, gives an
 estimate for $\theta_c$ and  
corrections to scaling should be observed as deviations from this power
 law decay of $C(t,T)$.   
Let us notice, however, that a precise measurement of $T_c$ can be only 
obtained for very large $\tau$. From dynamical scaling,  evolution 
up to a finite time $t_f$ corresponds to equilibration up to a size 
$L(t_f)$, and the accuracy in the determination of $T_c$ is then of the same 
order of FSS analysis on system up to size $L(t_f)$.  
In the following we present results involving not very large simulation 
times ($\sim 12 h$ of cpu time for each temperature), that, however,  
are sufficient to identify the critical temperature with a reasonable
accuracy. For each given system and each temperature we consider about
$1000$ independent realizations. 
   
Dynamical evolution is obtained via standard Monte Carlo
 simulations and $X(T,\tau)$ is obtained after the integration of
 $C(t,T)$ with a time-step of single spin update. 
In all cases we always take different sample sizes $L$
 in order to check that no finite size effects are present.

\begin{figure} 
   \centering 
  \rotatebox{0}{\resizebox{.4\textwidth}{!}{\includegraphics[clip=true]{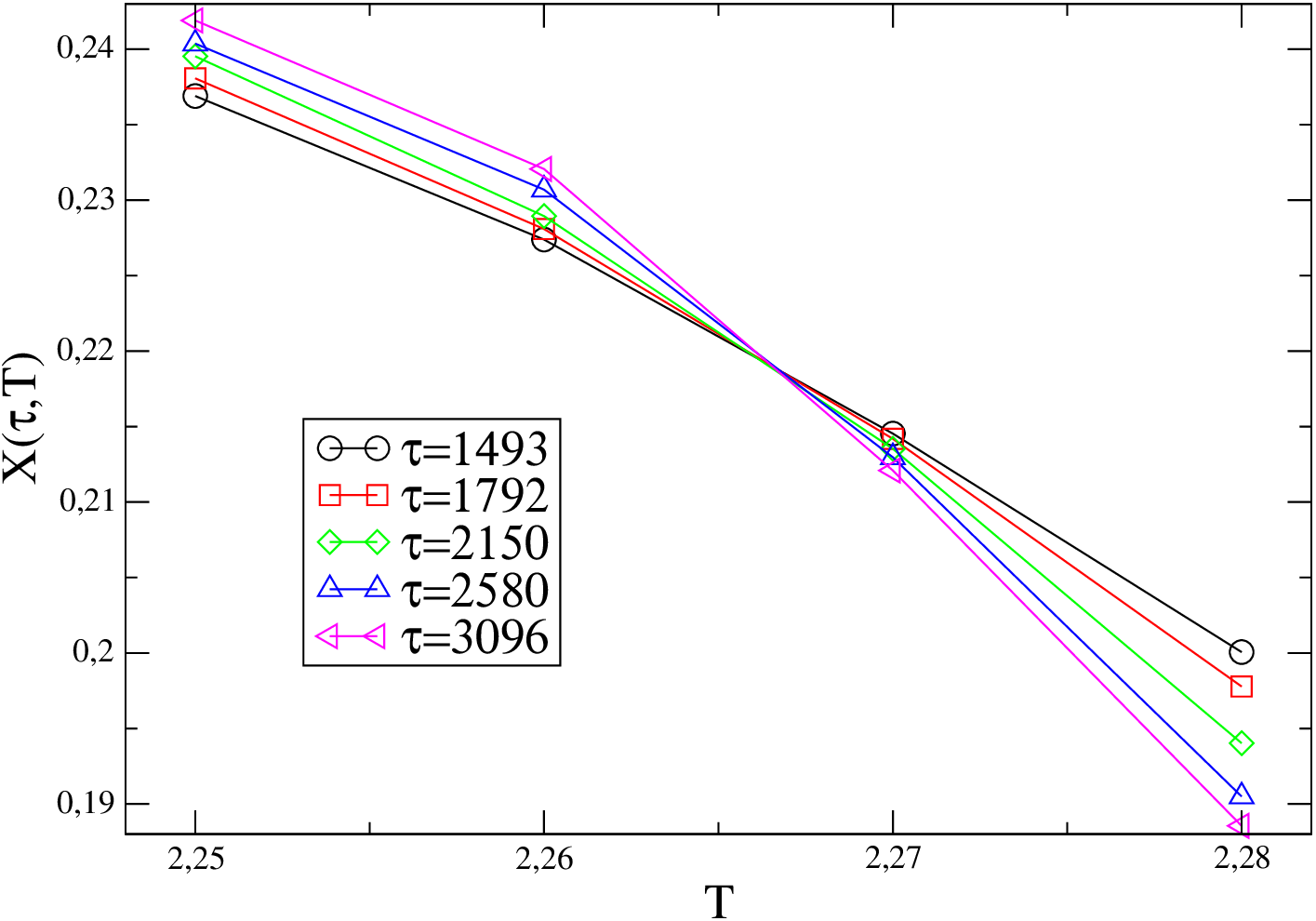}}} 

  \vspace{1.0cm} 

  \rotatebox{0}{\resizebox{.4\textwidth}{!}{\includegraphics{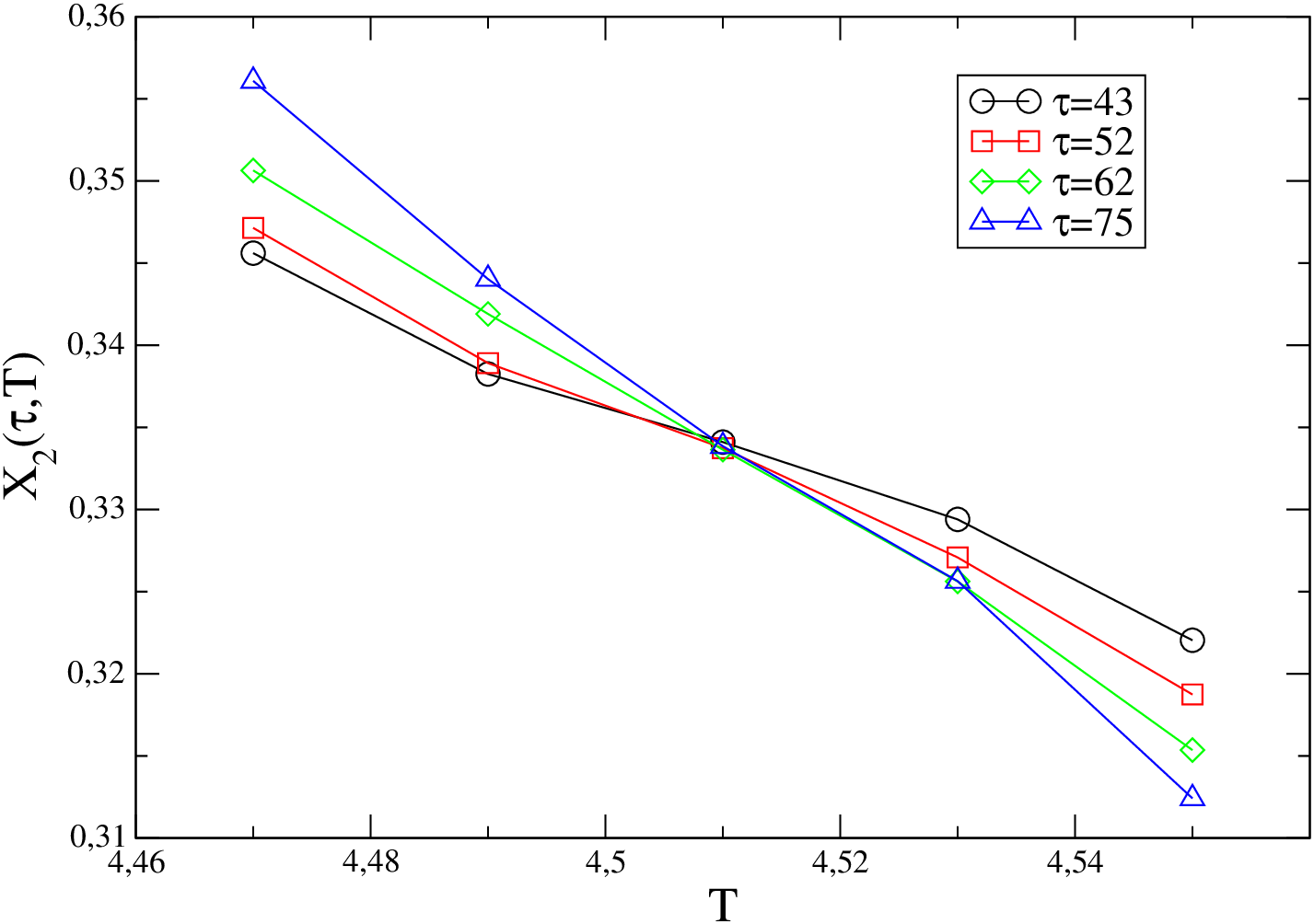}}} 
   \caption{(Color online). $X(\tau,T)$ and $X_2(\tau,T)$ are plotted versus $T$ 
     for different values of $\tau$ for the Ising model in $d=2$ and $d=3$, respectively.  
     Data cross around the known critical temperatures $T_c=2.269$ and $T_c=4.511$.
     In $d=2$, $\theta\sim 0.625$ for $T<T_c$ and $\theta_c\sim 0.74$, while in $d=3$, 
     $\theta\sim 1.3$ for $T<T_c$ and $\theta_c\sim 1.4$. In the inset of the upper panel
     the quantity $X(\tau,T)$ is plotted over $\tau$ for different temperatures.
     In the numerical simulations we consider systems of $N=L^d$ spins,  
     with $L=400$ for $d=2$ and $L=128$ for $d=3$. } 
\label{isi} 
\end{figure}

Let us begin by checking our method in cases where the critical temperature is well known. 
In particular, we start by considering the  Ising model with Hamiltonian 
$\cal{H}=-\sum_{\langle ij\rangle}J_{i j}s_i s_j$ 
and ferromagnetic coupling $J_{i j}=J$  in two and three
dimensions. 
In these cases  
$T_c\simeq 2.269J$ and $T_c\simeq 4.5115J$ are, respectively, analytically  
and numerically known~\cite{blote}. In the following we always take $J=1$ for simplicity. 
The behavior of $C(t,T)$ can be obtained from 
general arguments~\cite{std,huse,calabrese}, giving $C(t,T)\sim t^{-\theta_c}e^{-t/t_{eq}(T)}$ 
for $T\ge T_c$, where $\theta_c$ can be related to static and dynamic critical exponents, 
and $C(t,T)\sim t^{-\theta}$ with $\theta<\theta_c$ for $T<T_c$.  
Then one has, for large $\tau$ and $\theta_c<1$
\begin{equation}
X(\tau,T) \to \left \{
\begin{array}{ll} t_{eq}(T)/\tau   & \textrm{for}~ T\gtrsim T_c \\
  (1-\theta_c)/(2-\theta_c) & \textrm{for}~ T=T_c \\
X_0=(1-\theta)/(2-\theta) & \textrm{for}~ T<T_c.  
\end{array} \right. 
\label{Xinf}
\end{equation}
More precisely, for $T<T_c$, the dynamics is initially attracted by the critical
point at $T_c$~\cite{bray} and then converges, for large $\tau$, to
$X_0>X(\tau,T_c)$.
This implies that
$X(\tau,t)$ diminishes by increasing $\tau$ for $T>T_c$, grows until
it converges to $X_0$ when $T < T_c$, and is $\tau$ independent at
$T_c$.
In the upper panel of Fig.~\ref{isi} the quantity $X(\tau,T)$ is plotted over the temperature 
for different values of $\tau$, for the two-dimensional Ising model evolving via Glauber dynamics. 
One clearly observes that the curves intersect in a narrow region giving  
$T_c=2.268 \pm 0.002$ in agreement with the analytical result. 

\begin{figure} 
   \centering 
  \rotatebox{0}{\resizebox{.4\textwidth}{!}{\includegraphics[clip=true]{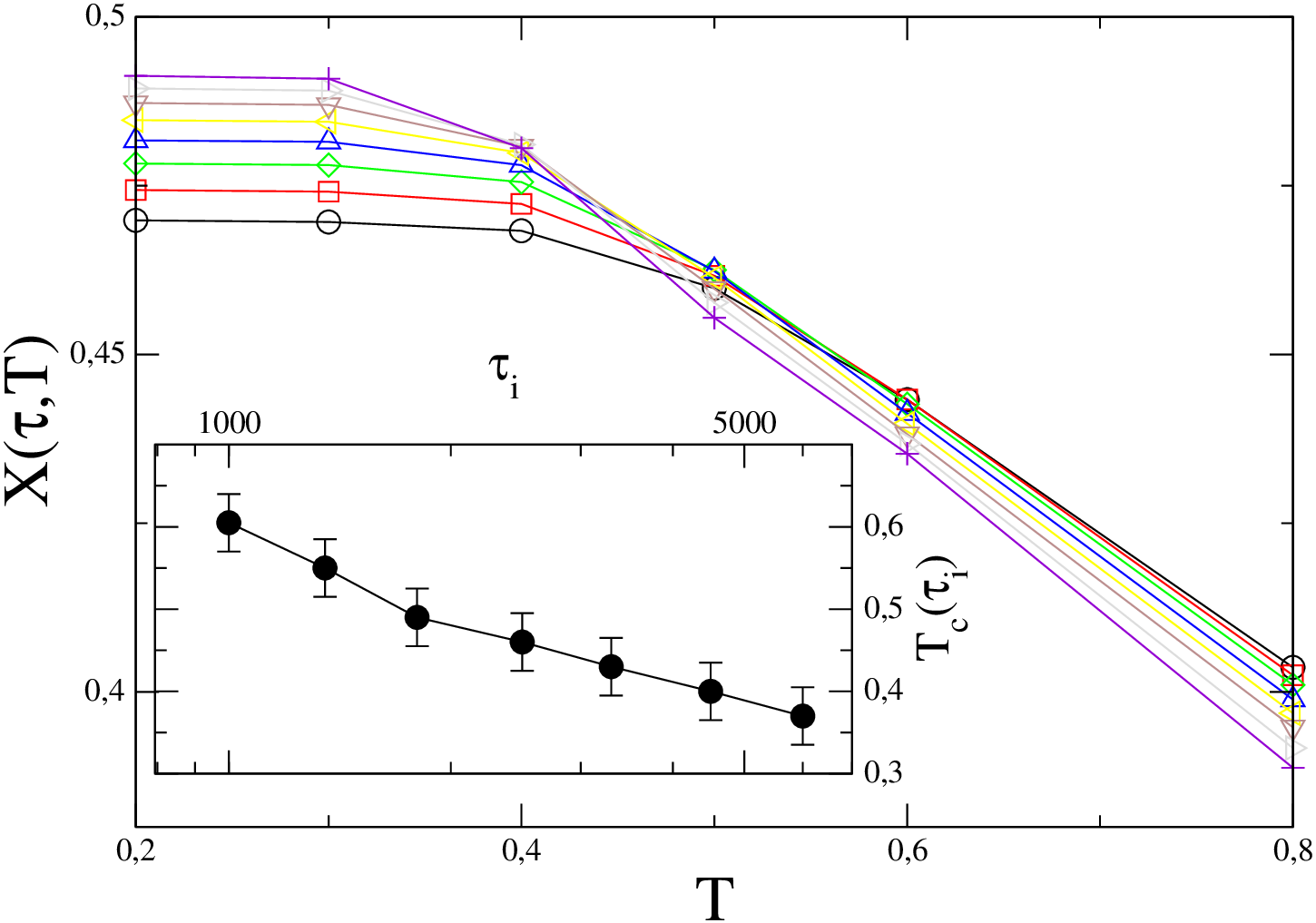}}} 
   \caption{(Color online). $X(\tau,T)$  vs $T$ 
     for different values of $\tau$ for the EA model in $d=2$. In the
     inset $T_c(\tau_i)$ is plotted vs  $\tau_i$ supporting the
     absence of a finite $T_c$.}
\label{ead2} 
 \vspace{1.0cm} 
\end{figure}

In the case of $d=3$, since $\theta_c\sim 1.4>1$~\cite{lambdaising3d},  
the integral $\int_0^\tau dt~t^{-\theta_c}$ diverges, making $X(\tau,T)$ useless for 
extracting critical behaviors. One then can overcome 
this problem considering the ``second moment''   
$X_2(\tau,T)=\frac{1}{\tau}\frac{\int _0^\tau dt~t^2 C(t,T)}{\int _0^\tau dt~tC(t,T)}$, 
that is expected to converge to $(2-\theta_c)/(3-\theta_c)$ for the critical quench.  
In the lower panel of Fig.~\ref{isi} the quantity $X_2(\tau,T)$ is plotted versus the temperature 
for different values of $\tau$. 
One again clearly observes that the curves intersect in a narrow region giving  
$T_c=4.510\pm0.003$, in agreement with previous numerical results. 

\begin{figure*}[!tb] 

\includegraphics[width=0.9\columnwidth,clip=true]{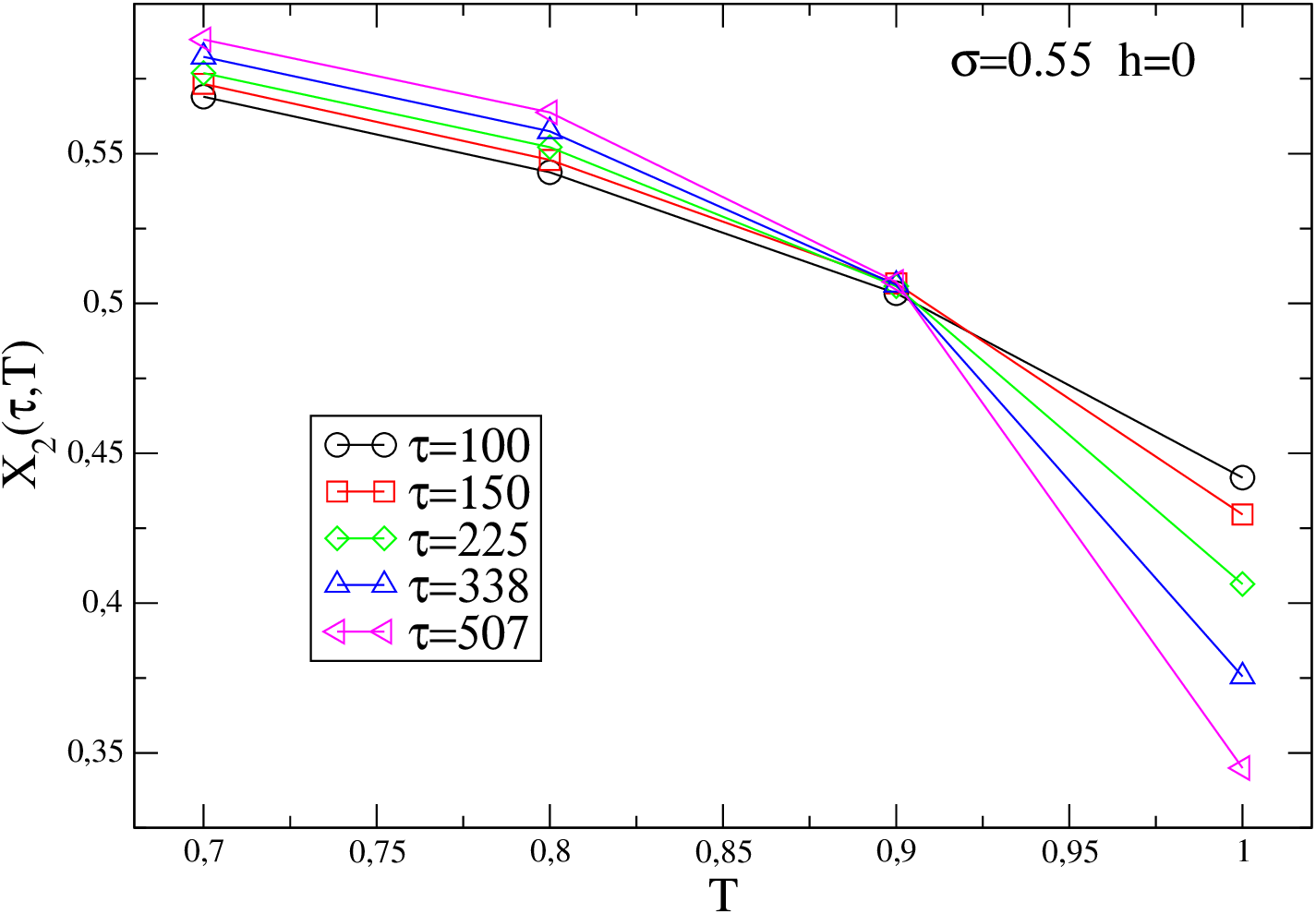}\qquad\qquad 
\hspace*{-1.0cm} 
\includegraphics[width=0.9\columnwidth]{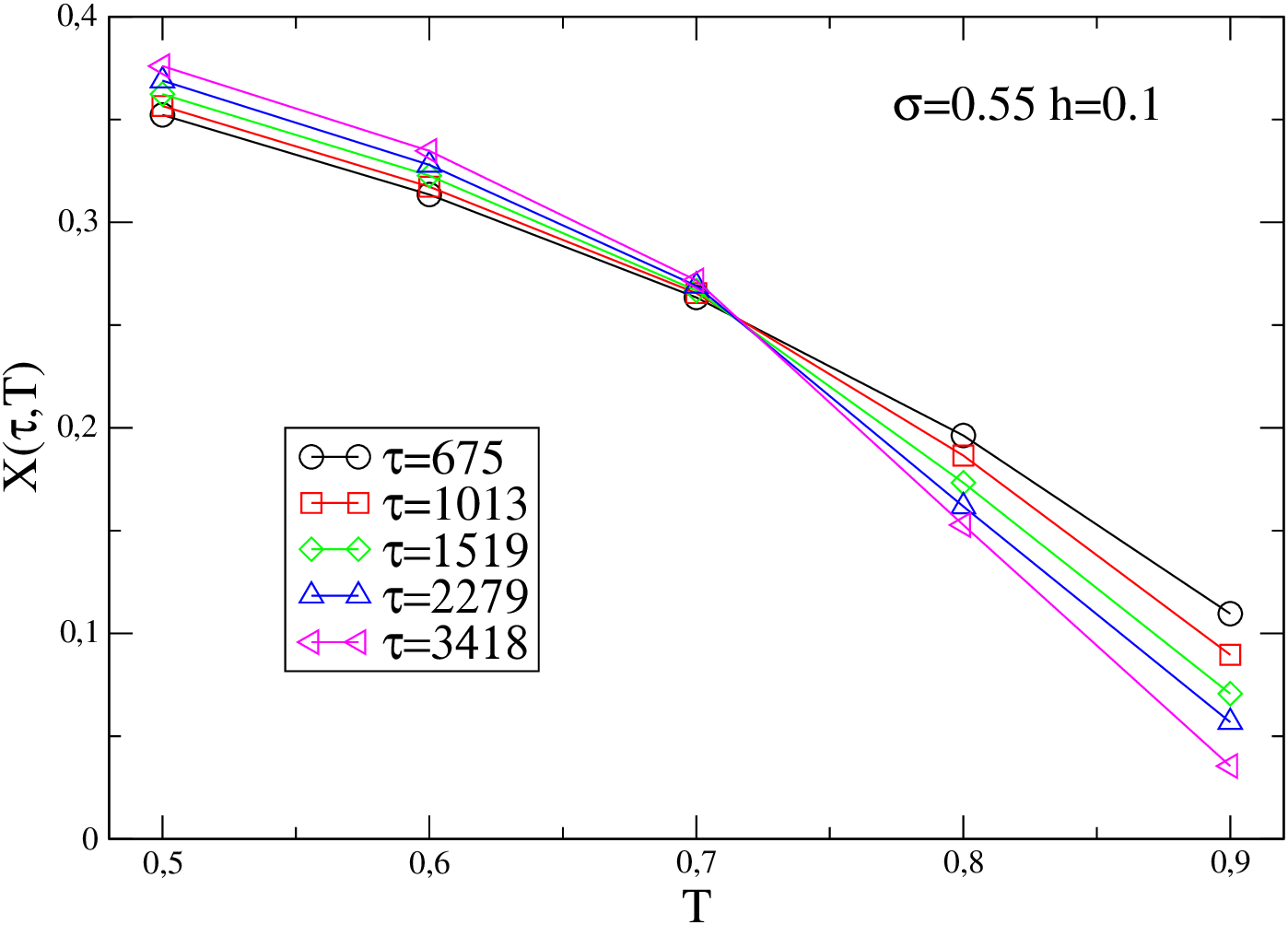}
\caption{(Color online). The quantities $X_2(\tau,T)$ and $X(\tau,T)$ are plotted 
versus the temperature for different values of $\tau$, in the one-dimensional Ising spin glass with power-law
interactions, for $\sigma=0.55$ and $h=0,0.1$. In all cases the curves cross, indicating the presence 
of a finite critical temperature.} 
\label{results1} 
\end{figure*} 

\begin{figure*}[!tb] 
\vspace*{0.8cm} 

\includegraphics[width=0.9\columnwidth]{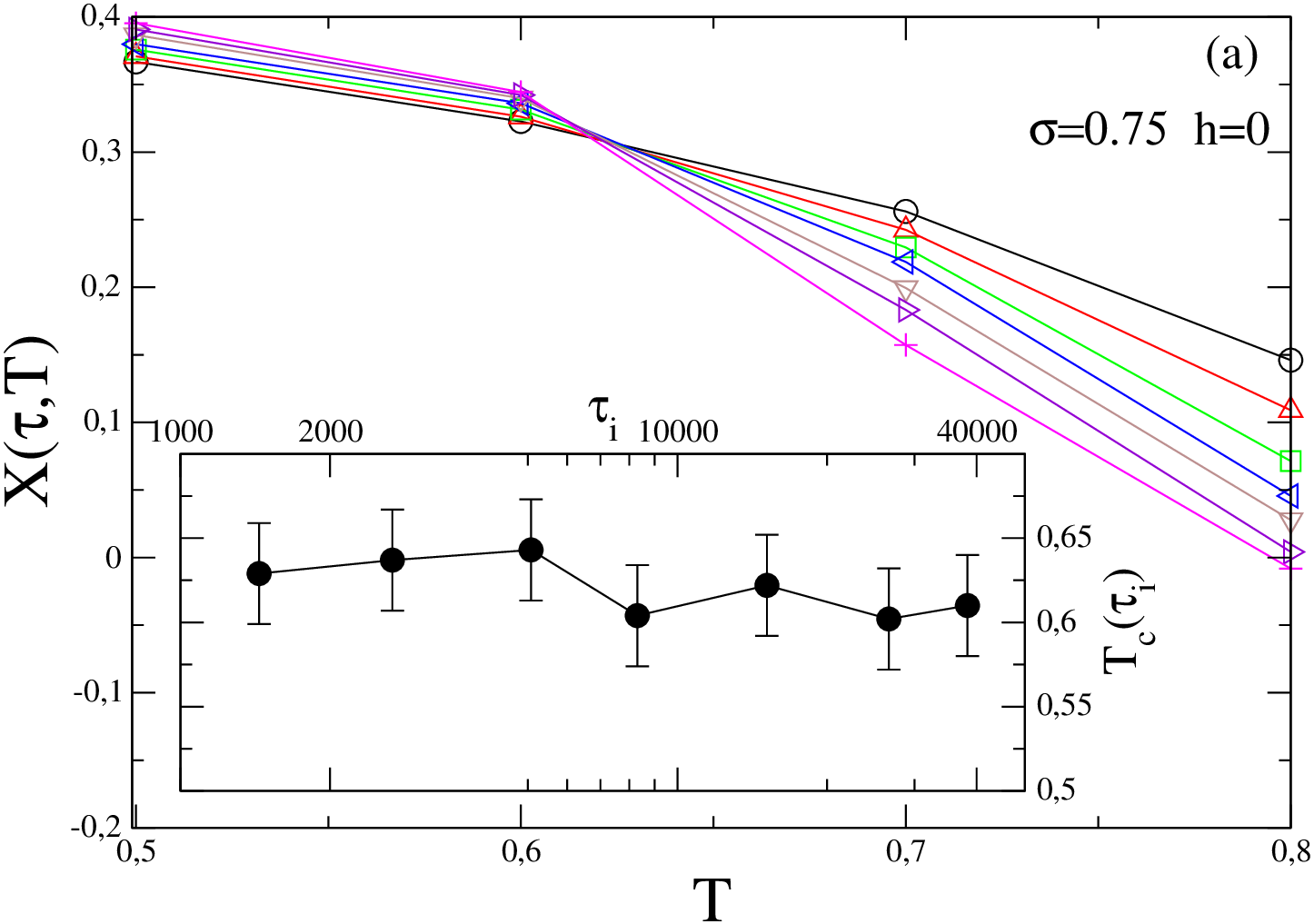}\qquad\qquad 
\hspace*{-1.0cm} 
\includegraphics[width=0.9\columnwidth]{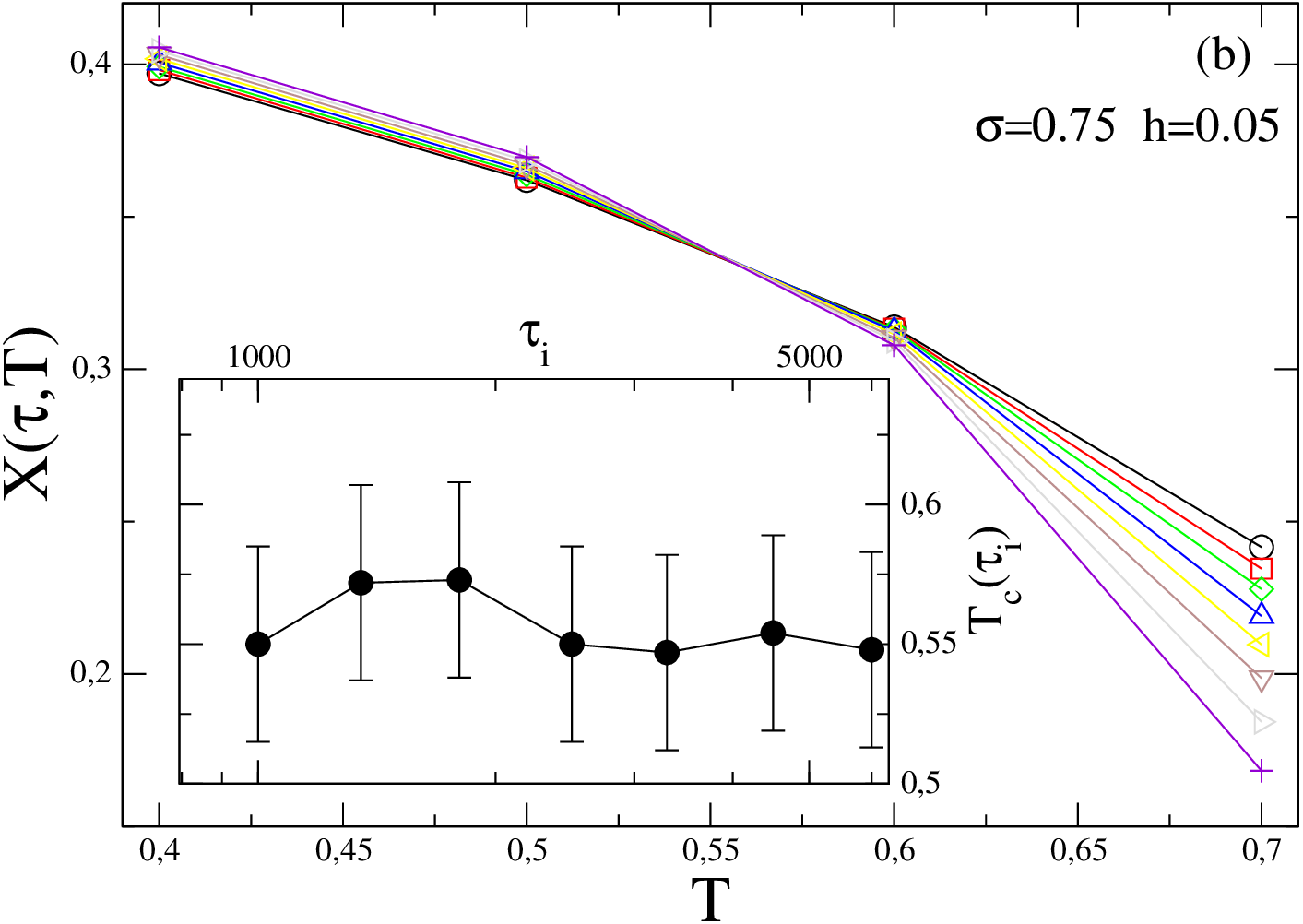} 

\vspace*{1.0cm} 

\includegraphics[width=0.9\columnwidth]{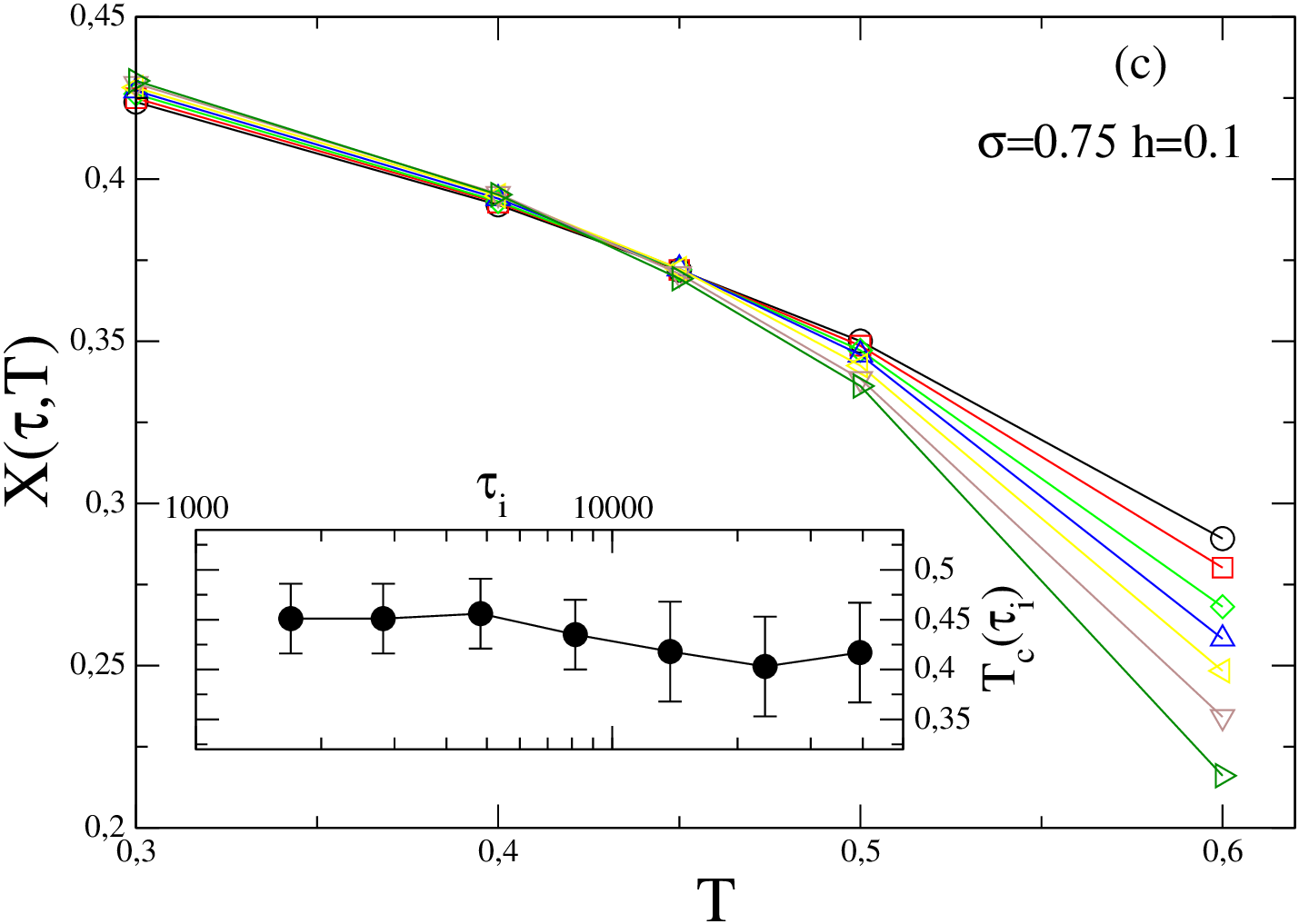}\qquad\qquad 
\hspace*{-1.0cm} 
\includegraphics[width=0.9\columnwidth]{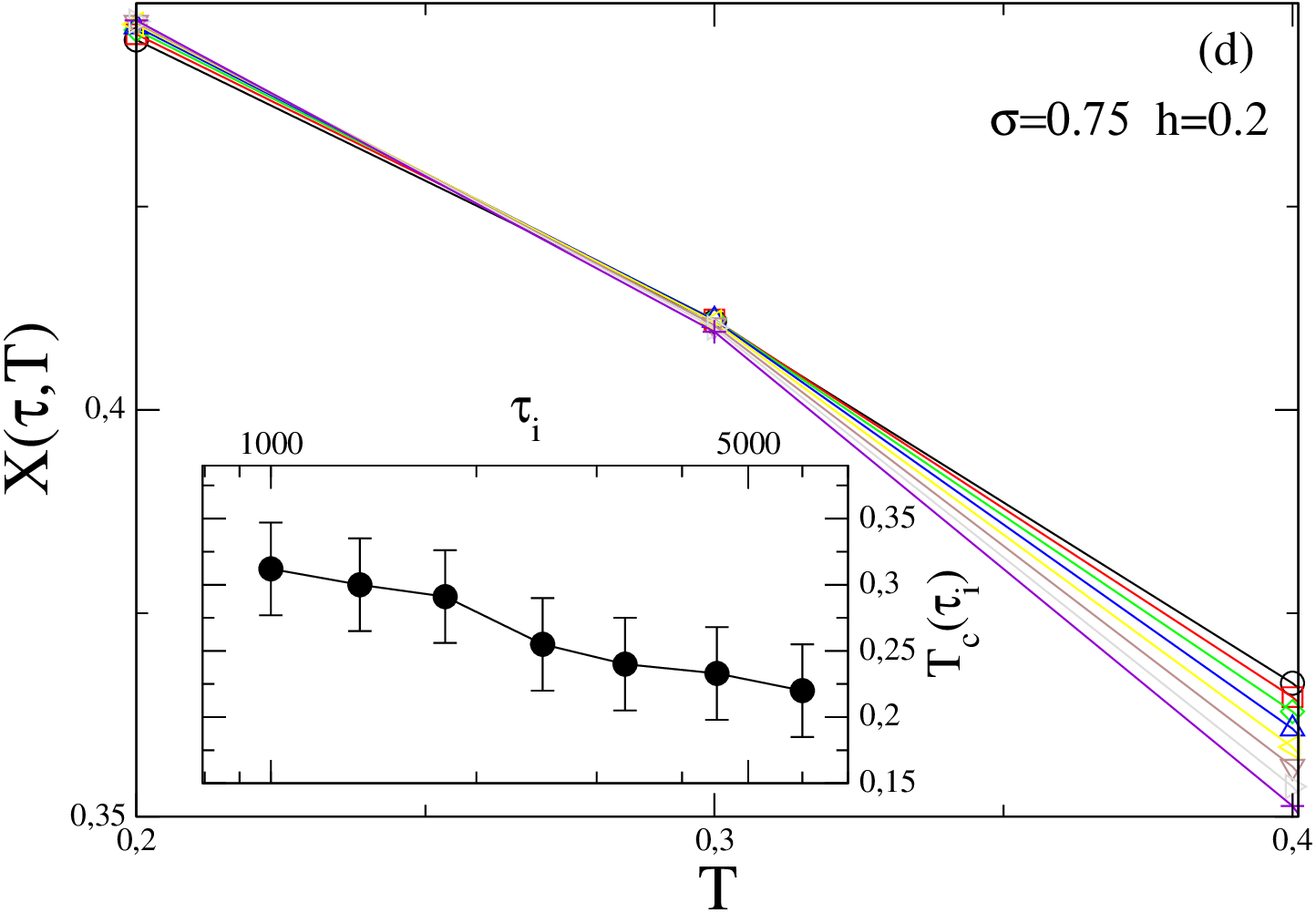} 
\caption{(Color online). The quantities $X(\tau,T)$ is plotted 
over the temperature in the one-dimensional Ising spin glass with power-law
interactions, for $\sigma=0.55$ and $h=0,0.05,0.1,0.2$. 
Curves correspond to different $\tau_i$ obtained by the relation
$\tau_{i+1}=1.35\tau_i$ and $\tau_1=1000$. In the insets the plot
$T_c(\tau_i)$ vs $\tau_i$ gives a finite critical temperature for
$h=0,0.05,0.1$ and suggests $T_c=0$ for $h=0.2$. } 

\label{results2} 
\end{figure*} 

As intermediate case between ferromagnets and spin glass systems, we  have
also considered the diluted Ising model. In this model the spin coupling
$J_{i j}$ is chosen to be $J$ with probability $p$ and to be $0$ with
probability $1-p$. Our results substantially agree with previous
accurate estimates of $T_c$ in Ref.~\cite{Ballesteros2}, for different choices of
the parameter $p$. 
As a further test, we have considered the $d=2$ Edwards-Anderson model,
where previous studies clearly indicate the absence of a phase
transition at finite temperature \cite{ead2ref1,ead2ref2,ead2ref3}. More precisely, we
have investigated  systems
with $N=400^2$ spins and bimodal couplings $J_{ij}=\pm1$ with equal
probability. Results, plotted in fig.\ref{ead2}, show that curves
corresponding to different $\tau$ do not intersect in the same point
but the intersection points move towards the left by increasing
$\tau$. More precisely, in the inset of fig.{\ref{ead2}}, we plot the
``critical temperature'' $T_c(\tau_i)$ identified from  the  intersection between
the curves for $X(\tau_i,T)$ and $X(\tau_{i+1},T)$. We
chose $\tau_{i+1}=\alpha \tau_i$ with $\tau_1=1000$ and
$\alpha=1.35$. The inset clearly shows that $T_c(\tau_i)$ decreases for
increasing $\tau_i$, indicating an asymptotic convergence $T_i \to 0$
at large times. This result supports the absence of a finite critical
temperature.  
The same analysis performed for the other cases where a finite
$T_c$ has been identified, gives a $T_c(\tau_i)$ fluctuating around $T_c$.

Next we consider the case of the one-dimensional Ising spin glass with power-law decaying
interactions~\cite{power.law}. 
The system is defined by the Hamiltonian 
${\cal H}=-\sum_{i,j}J_{ij}s_is_j-h\sum_is_i$, 
where the site
 $i$ belongs to a ring of length $L$ and $h$ is a magnetic field. 
The sum is over all  spins of the ring 
and $J_{ij}=c(\sigma)\epsilon_{ij}/r^\sigma_{ij}$, 
where $\epsilon_{ij}$ are chosen according to a Gaussian distribution with zero mean and standard deviation unity. 
The constant $c(\sigma)$ is chosen to give a mean field transition temperature 
$T_c^{MF}=1$, namely 

\begin{equation} 
1=(T_c^{MF})^2=\sum_{j\ne i}[J_{ij}^2]_{av}=c(\sigma)\sum_{j\ne i}\frac{1}{r_{ij}^{2\sigma}}, 
\label{TcMF} 
\end{equation} 
where $[\ldots]_{av}$ denotes an average over disorder. The distance between two spins on the ring 
in terms of $L$ is $r_{ij}=(L/\pi)\sin(\pi|i-j|/L)$. 
By varying the strength of the interaction through the parameter $\sigma$, this model 
shows different behaviors~\cite{power.law}. In particular, for $\sigma=0$, taking $c(\sigma)\sim 1/\sqrt{N}$, one recovers the  
Sherrington-Kirkpatrick model~\cite{SK}. 
For $\sigma\in [1/2,1]$ the system shows a finite critical temperature, 
with a mean field-like region for $\sigma\in[1/2,2/3]$ and a non-mean field 
region for $\sigma\in(2/3,1]$.  
In recent years, this model and its diluted version have been widely investigated in the 
literature~\cite{leuzzi1,katzgraber,katz,leuzzi2,katz2}, 
focusing on the identification 
of a transition in presence of an external field, namely on the identification   
of the AT line.  
In these studies, $T_c$ for different values of $\sigma$,  
both with and without the external field, has been measured with 
FSS analysis~\cite{leuzzi1,katzgraber,katz,leuzzi2,katz2}. Contradictory results have been obtained  
in the non mean field-like region in presence of the external
perturbation. When $h \ne 0$, indeed,  
Leuzzi {\it et al.}~\cite{leuzzi2} find a finite critical temperature, whereas no transition  
has been observed by Katzgraber \& Young~\cite{katz,katz2}. 
\begin{figure} 
   \centering 

  \rotatebox{0}{\resizebox{.4\textwidth}{!}{\includegraphics[clip=true]{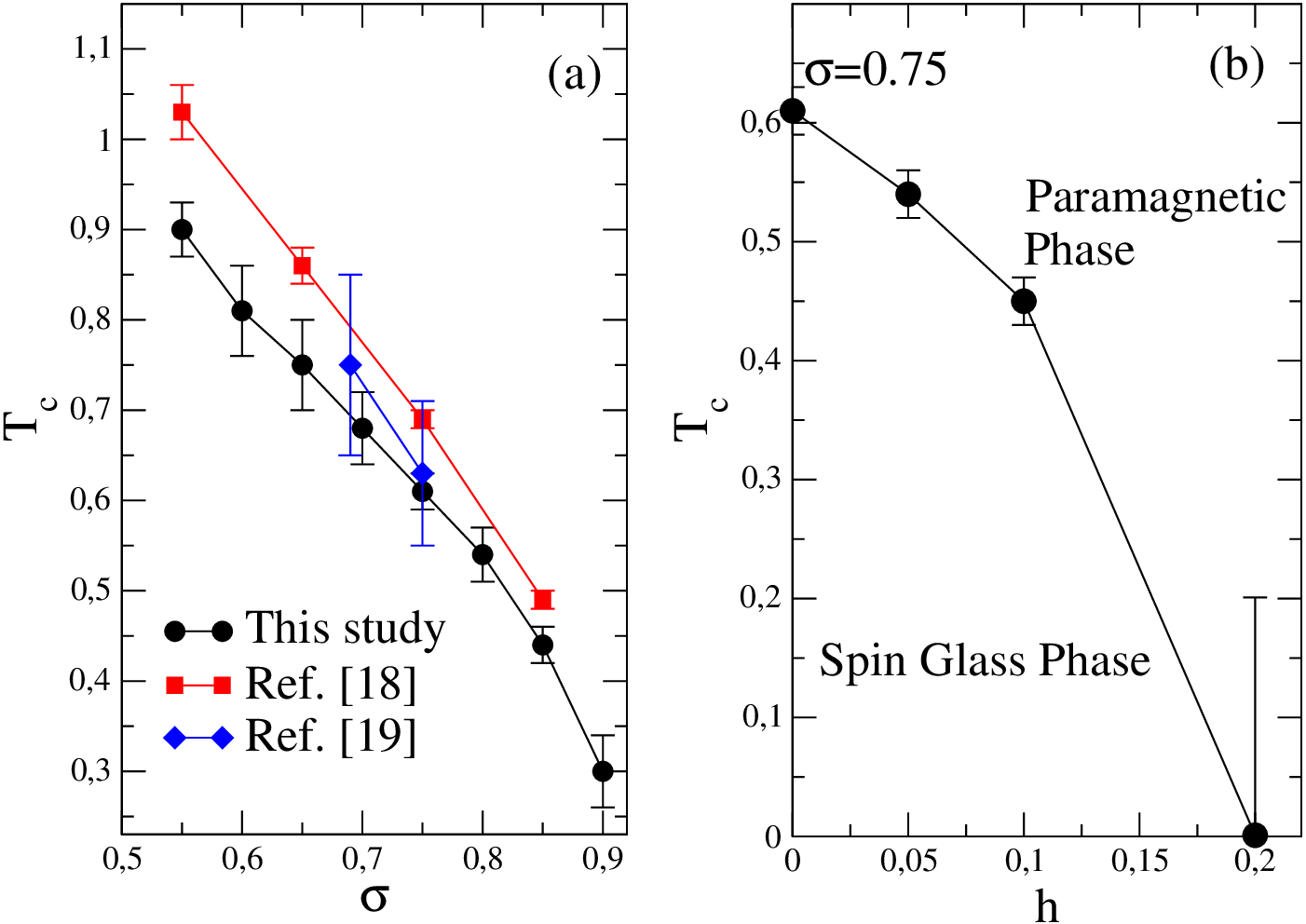}}}  
   
   \caption{(Color online). Panel (a): the critical temperatures of the  
     power law one-dimensional Ising spin glass measured in Refs.~\cite{katz},~\cite{leuzzi1} 
     and in this study are plotted for several values of $\sigma$ in the absence of external field. 
     Panel (b): the $T-h$ phase diagram is reported for the same model with $\sigma=0.75$.} 
\label{sigmaT} 
\end{figure}  
We turn to consider our results.
In all our simulations we consider systems with  
$N=1024$ spins and final times up to 10000 Monte Carlo
  steps.
 We have explicitly checked that no finite size effects are present. 
We first discuss results for different values of $\sigma\in[0.5,0.9]$ without external field. 
In particular, in the left panels of Figs.~\ref{results1} and ~\ref{results2}, we plot
$X_2(\tau,T)$ for $\sigma=0.55$, since $\theta_c =1.09 \pm 0.02>1$, and $X(\tau,T)$ 
for $\sigma=0.75$ versus $T$ for different $\tau$. 
Curves clearly show intersection points, giving $T_c=0.90\pm0.02$  
and $T_c=0.61\pm0.02$ for $\sigma=0.55$ 
and for $\sigma=0.75$, respectively. 
The critical temperature obtained from the same analysis for other values of $\sigma$ 
are reported in the left panel of Fig.~\ref{sigmaT}, where previous results~\cite{katz,leuzzi1} are also shown. 
We find that $T_c$ is a monotonously decreasing function of $\sigma$, consistent 
with a linear decay $T_C\sim 1-1.6(\sigma-0.5)$, for $\sigma\in[0.5,0.8]$ and
a faster decay for larger $\sigma$. The monotonic decreasing behavior
of $T_c$  with increasing $\sigma$ 
is expected, since larger values of $\sigma$ correspond to shorter
interaction ranges. 
Conversely, the results of Ref.~\cite{katz} 
show a decreasing linear behavior
for $\sigma\ge 0.55$, but are expected to manifest a 
non-monotonic behavior approaching  $\sigma=0.5$, where the mean field
value $T_c=1$ is imposed by Eq.~(\ref{TcMF}). 
We wish to notice that, for every choice of $\sigma$, we
always obtain  a value of $T_c$ significantly smaller 
than the one obtained in Ref.~\cite{katz}. 
The value of $T_c$ estimated in Ref.~\cite{leuzzi1}  are  intermediate
between our results and those of Ref.~\cite{katz}, but due to the large
error bar, are compatible with both findings.

Differences between the results of our method and those of
Ref.~\cite{katz} become more pronounced  when  $h\ne 0$. 
More precisely, in the mean field region, for $\sigma=0.55$ and $h=0.1$ we 
find $T_c=0.71\pm0.03$, a value still smaller than that of Ref.~\cite{katz},
$T_c=0.96\pm 0.02$. Conversely, an opposite trend is obtained 
in the non mean field region $\sigma>2/3$, where we always find a
$T_c>0$, while no transition was obtained in Ref.~\cite{katz}.      
Let us stress that we can consider  
a uniform field $h_i=h$, 
whereas FSS analysis imposes the application of a spatially
decorrelated (random) field. 
  
In particular, we focus on $\sigma=0.75$ and three values of $h$,
$h=0.05,0.1,0.2$. For $h=0$ and $h=0.1$, we consider longer
  simulations up to 50000 Monte Carlo steps and N=2048 in order to
  avoid finite size effects.
Results for $X(T,\tau)$ are plotted in the main
panels of Fig.~\ref{results2}. The behavior of $T_c(\tau_i)$ vs
$\tau_i$ is plotted in the insets. Fig.~\ref{results2}b
  and Fig.~\ref{results2}c show that curves intersect at a finite
  temperature indicating 
the existence of a phase transition 
in presence of an external field. We notice that the curves spreading at low
temperature is less pronounced than for smaller $\sigma$ and other
models. This can be attributed to the smaller value of
$\theta_c$. Indeed, from Eq.~(\ref{Xinf}), at fixed difference
$\theta_c-\theta$  one has that $X_0-X(\tau,T_c)$ is a decreasing
function of $\theta_c$.  
The measured values
$T_c=0.56 \pm 0.01$ and $T_c=0.43 \pm 0.01$, for $h=0.05$ and
$h=0.1 $, respectively,  are consistent with the expected trend of a
decreasing $T_c$  for increasing $h$. The inset of
Fig.~\ref{results2}d gives a non-constant $T_c(\tau_i)$ that decreases
at small $\tau_i$ and tends to flattens only for the largest
$\tau_i$. The above trend suggests the absence of a finite $T_c$ for
$h=0.02$ but does not exclude that $T_c(\tau_i)$ asymptotically
converges to  $T_c \in (0,0.2)$ at large times.         
In the right panel of Fig.~\ref{sigmaT} we plot the $T-h$ 
phase diagram showing the existence of the AT line, 
separating the spin glass phase from the paramagnetic one. 

In order to obtain more insights on the behavior of
  disordered systems, the same procedure can be carried out replacing
  $C(t)$ in Eq.(\ref{FTS}) by a multi-spin correlation
  function. Nevertheless, the non linear susceptibility 
$\chi_4(t,t_w)=\sum_{i,j} \langle
  s_i(t)s_j(t)s_i(t_w)s_j(t_w)\rangle$ usually considered in the
  investigation of disordered systems \cite{franz}, is not properly suitable for
  this kind of study. Indeed, 
$\chi_4 (t, t_w )$ encodes the typical length scale $L(t)$ of spatial
correlation, and is expected to grow in time until a limit
value that depends on $\xi(T )$ or $L(t_w )$ \cite{noi}.
 More precisely,
for $L(t_w ) < \xi(T )$,$ X(\tau, T )$  would be not 
affected by $\xi(T)$ and, therefore, it would not be able to identify eventual
divergences of $\xi(T )$. A way to overcome the above 
difficulty is to consider other multi-spin correlations, such as
the second-order susceptibility considered in Ref.~\cite{noi2}. This
quantity is intimately related to $\chi_4 (t, t_w)$ and converges to
an asymptotic value controlled by $\xi(T )$, independently of
$t_w$ . This quantity can be used in the described procedure.
Difficulties, in this case, are related to huge fluctuations of this
second order susceptibility that make the numerical evaluating very
time-demanding.

In conclusion, we have introduced an dimensionless quantity $X(\tau,T)$
that is expected to become time independent at the critical
temperature $T_c$. This allows one to identify $T_c$ from the intersection
of curves $X(\tau,T)$ for fixed $\tau$ and different $T$.
The method has been tested in models where there
exist accurate estimates of $T_c$, as the Ising ferromagnet
and its diluted version. The study of the two-dimensional EA model 
confirms the absence of a transition at finite temperature. 
The method has been then applied to the one-dimensional Ising spin
glass with power law decaying interactions for different choices of
$\sigma$ and $h$. Results for $h=0$ give $T_c$ values always smaller
than those obtained by static FFS methods in Ref.~\cite{katz}, with differences that are
larger for smaller $\sigma$. In particular for $\sigma=0.55$ we obtain
$T_c=0.90\pm0.02$, a value smaller than the one of Ref.\cite{katz}, where
$T_c=1.03\pm0.03$, unexpectedly above the mean-field value $T_c=1$. The study
in presence of a finite perturbation $h>0$ indicates the existence of
a finite critical temperature also in the non mean-field-like region   
$\sigma>2/3$, in agreement with a  replica symmetry breaking scenario.

\acknowledgments
We thank Marco Zannetti, Federico Corberi, Luca Leuzzi and Lucilla
de Arcangelis  for useful discussions and observations.  
A.Sarracino acknowledges financial support from PRIN 2007 JHLPEZ.

\end{document}